\documentclass[10pt,a4paper]{article}

\usepackage[margin=2cm]{geometry}
\usepackage{amsfonts}
\usepackage{graphicx}
\usepackage{subfig}
\usepackage{cite}
\usepackage{caption}
\usepackage{amsmath}
\usepackage{amssymb}
\usepackage{color}
\usepackage{authblk}
\usepackage[utf8]{inputenc}

\title{\textbf{Kerr black holes with synchronised scalar hair \\ and higher azimuthal harmonic index}}
\author[1]{Jorge F. M. Delgado\footnote{jorgedelgado@ua.pt}}
\author[2]{Carlos A. R. Herdeiro\footnote{carlosherdeiro@tecnico.ulisboa.pt}} \author[1]{Eugen Radu\footnote{eugen.radu@ua	.pt}}
\affil[1]{\normalsize Departamento de Física da Universidade de Aveiro and

Center for Research and Development in Mathematics and Applications -- CIDMA

Campus de Santiago, 3810-183 Aveiro, Portugal \vspace{10pt}}

\affil[2]{Centro de Astrofísica e Gravitação -- CENTRA,

Departamento de Física, Instituto Superior Técnico -- IST, Universidade de Lisboa -- UL,

Avenida Rovisco Pais 1, 1049-001 Lisboa, Portugal }

\begin{document}

\date{March 2019}

\maketitle

\begin{abstract}
\normalsize
Kerr black holes with synchronised scalar hair and azimuthal harmonic index $m>1$ are constructed and studied. The corresponding domain of existence has a broader frequency range than the fundamental $m=1$ family; moreover, larger ADM masses, $M$ and angular momenta $J$ are allowed. Amongst other salient features, non-uniqueness of solutions for fixed global quantities is observed: solutions with the same $M$ and $J$ co-exist, for consecutive values of $m$, and the ones with larger $m$ are always entropically favoured. Our analysis demonstrates, moreover, the qualitative universality of various features observed for $m=1$ solutions, such as the shape of the domain of existence, the typology of ergo-regions, and the horizon geometry, which is studied through its isometric embedding in Euclidean 3-space.
\end{abstract}


\section{Introduction}
Kerr black holes (BHs) with synchronised scalar hair~\cite{Herdeiro:2014goa}  are a counterexample to the no-hair conjecture~\cite{Ruffini:1971bza} -- see~\cite{Herdeiro:2015waa,Sotiriou:2015pka,Volkov:2016ehx} for reviews -- occurring in a simple and physically sound model: Einstein-(complex and massive-)Klein-Gordon theory.  Many related solutions, relying on a similar synchronisation mechanism, have been found in the last few years, in different setups and approximations. An incomplete list of references, including also various studies of physical properties, is~\cite{Dias:2011at,Hod:2012px,Barranco:2012qs,Hod:2013zza,Herdeiro:2014ima, Herdeiro:2014jaa, Hod:2014baa,Benone:2014ssa,Herdeiro:2015gia, Smolic:2015txa, Kleihaus:2015iea, Cunha:2015yba, Herdeiro:2015tia, Iizuka:2015vsa, Herdeiro:2015kha, Wilson-Gerow:2015esa, Chodosh:2015oma, Hod:2016yxg, Herdeiro:2016gxs, Herdeiro:2016tmi, Huang:2016qnk, Cardoso:2016ryw, Cunha:2016bpi, Brihaye:2016vkv, Vincent:2016sjq, Ni:2016rhz, Delgado:2016jxq, Bernard:2016wqo, Smolic:2016dmh, Cunha:2016bjh, Sakalli:2016xoa, Hod:2016lgi, Franchini:2016yvq, Hod:2017kpt, Barranco:2017aes, Cunha:2017eoe, Grover:2017mhm, Huang:2017whw, Barjasic:2017oka, East:2017ovw, Dolan:2017otg, Herdeiro:2017phl, Ferreira:2017pth, Brihaye:2017wqa, Palenzuela:2017kcg, Hod:2018pri, Ferreira:2017cta, Collodel:2017end, Ganchev:2017uuo, Herdeiro:2017oyt, Cunha:2018acu, Degollado:2018ypf, Baumann:2018vus, Brihaye:2018mlv, Delgado:2018khf,Peng:2018wtq,Herdeiro:2018daq, Wang:2018xhw,Gimeno-Soler:2018pjd,Garcia:2018sjh,Peng:2019waa}.

These hairy BH solutions have a relation with the physical phenomenon of superradiance~\cite{Brito:2015oca}, from which they can form dynamically from the Kerr solution~\cite{East:2017ovw, Dolan:2017otg, Herdeiro:2017phl} - see also~\cite{Ganchev:2017uuo,Degollado:2018ypf} for a discussion on the metastability of these solutions against superradiance. They also reduce to Kerr BHs and boson stars~\cite{Schunck:2003kk,Liebling:2012fv}, in appropriate limits. Boson stars are a sort of gravitating soliton interpreted as a Bose-Einstein condensate of an ultra-light scalar field,  that could be a dark matter candidate~\cite{Suarez:2013iw,Hui:2016ltb}. Moreover, the existence of the hairy BH solutions does not rely on particular choices of scalar field potentials that violate energy conditions, unlike other examples of asymptotically flat BHs with scalar hair, see $e.g.$~\cite{Nucamendi:1995ex,Cadoni:2015gfa}. Thus, besides the issue of the no-hair conjecture in BH physics, these hairy BHs contain different angles of interest. 

Kerr BHs with synchronised hair comprise a family that, besides the continuous parameters mass, angular momentum and Noether charge, is labelled by two discrete numbers: the azimuthal harmonic index of the scalar field $m\in \mathbb{Z}^+$ and its node number $n$. Most of the studies of the solutions have focused on the fundamental solutions, $n=0$, with the smallest value of $m=1$. Recently, excited solutions ($n\neq 0$) have also been constructed~\cite{Wang:2018xhw}. Solutions with $m>1$, on the other hand, have only been considered in the solitonic (boson star) limit~\cite{Grandclement:2014msa,Herdeiro:2015gia}, with the exception of the non-minimal model studied in~\cite{Kleihaus:2015iea}. The purpose of this work is to construct solutions with $m>1$ in the minimal, simplest model, and to study some of the basic physical properties of these new solutions. 

One motivation to study the higher $m$ solutions is that the superradiant instability of a given $m$ solution could drive it to migrate to an $m+1$ solution, in an asymptotic cascading process leading to $m\rightarrow \infty$~\cite{Dias:2011at}. This process is, likely, non-conservative, ejecting some energy and, especially, angular momentum towards infinity; but for particular solutions with a given $m$, if a neighbouring solution (in terms of global quantities) exists for $m+1$, the process could be approximately conservative. In fact, this approximate conservativeness has been observed in the transition from the Kerr BH (which corresponds to $m=0$) to the $m=1$ hairy solution in~\cite{East:2017ovw, Dolan:2017otg, Herdeiro:2017phl}. For this approximately conservative migration to be possible, the higher $m$ neighbouring solution would have to be entropically favoured. As we shall see herein, this is always the case: comparing solutions with consecutive values of $m$ with the same global quantities, the higher $m$ solution has a larger horizon area. 

Another motivation for studying this higher $m$ solutions is to assess the universality of some physical properties. For instance, it was observed in~\cite{Herdeiro:2014jaa} that, when scanning the domain of existence, these BHs exhibit a more diverse structure of ergo-regions than the standard one of the Kerr BH. The existence of these ergo-regions is at the origin of the superradiant instability. So, a natural question is if a similar structure is present for higher $m$. We shall see here that this is the case. Moreover, the horizon geometry of these hairy BHs has been recently studied in~\cite{Delgado:2018khf}, where it was found that the key property for deciding whether the horizon is embeddable in Euclidean 3-space is the horizon sphericity.  Again, we shall see that this is also the case for the higher $m$ solutions. Both these analyses provide evidence that the properties observed for $m=1$ solutions are universal throughout the whole discrete family labelled by $m$.

This paper is organised as follows. The model is presented in Section~\ref{section2}, together with some of the most relevant physical quantities for our study. The construction of the domain of existence of the $m=2,3$ solutions is presented in Section~\ref{section3}, where they are compared with the $m=1$ case. In Section~\ref{section4} the phase space is discussed and the entropy comparison shown. In Section~\ref{section5} other physical properties, in particular, the ergo regions and horizon geometry, are discussed. Section~\ref{section6} wraps up the paper with a discussion.

\section{The Model}
\label{section2}

Kerr BHs with synchronised scalar hair~\cite{Herdeiro:2014goa} are solutions of the Einstein-Klein-Gordon equations,
\begin{equation}
\label{eqmot}
R_{ab} - \frac{1}{2} g_{ab} R = \frac{8\pi G}{c^4} T_{ab} \ ,\hspace{15pt}  \hspace{15pt} \Box \Psi = \mu^2 \Psi \ ,
\end{equation}
where the energy-momentum tensor is
 $T_{ab} = 2\partial_{(a} \Psi^* \partial_{b)} \Psi - g_{ab}
 \left( \partial_c \Psi^* \partial^c \Psi + \mu^2 \Psi^* \Psi \right)$ and $\mu$ is the mass of the (complex) scalar field.\footnote{Henceforth we shall use units such that $G = c = \hbar = 1$.}
Such solutions represent a Kerr BH in equilibrium with a massive scalar field configuration and they were obtained numerically using the following \textit{ansatz},
\begin{equation}\label{Eq:NumericMetric}
ds^2 = - e^{2F_0} N dt^2 + e^{2F_1} \left( \frac{dr^2}{N} + r^2 d\theta^2 \right) + e^{2F_2} r^2 \sin^2 \theta \left( d\varphi - W dt \right)^2 \hspace{10pt}, \hspace{10pt} \Psi = \phi\ e^{i(m\varphi - \omega t)} \ ,
\end{equation}  
where $F_0, F_1, F_2, W$ and $\phi$ are \textit{ansatz} functions that only depend on $(r,\theta)$ coordinates, $\omega$ and $m = \pm 1, \pm 2, \dots$ are the angular frequency and azimuthal harmonic index of the scalar field, respectively, and $N \equiv 1 - {r_H}/{r}$, in which $r_H$  is the radial coordinate of the event horizon, which sits at $r=r_H=$ constant. An existence proof of these solutions was provided in~\cite{Chodosh:2015oma}.

The existence of these solutions relies on the so-called \textit{synchronization condition}. This condition can be interpreted as a synchronisation between the horizon angular velocity of the BH, $\Omega_H$, and the phase angular velocity of the scalar field, $\omega/m$, hence justifying its name:
\begin{equation}
\omega = m \Omega_H \ .
\end{equation} 

Our goal here is to study solutions with larger azimuthal harmonic index, namely $m=2,3$, as all previous studies for the model~\eqref{eqmot} have focused on $m=1$ solutions.

\subsection{Physical Quantities}

Most physical quantities of interest can be obtained, as in previous works, through the metric functions at the event horizon or spacial infinity. At the horizon, one computes the Hawking temperature, $T_H$, and horizon area, $A_H$, as~\cite{Herdeiro:2014goa},
\begin{equation}
T_H = \frac{1}{4\pi r_H} e^{\left(F_0 - F_1 \right)|_{r_H}}\ , \hspace{10pt}  \hspace{10pt} A_H = 2\pi r_H^2 \int_0^\pi d\theta \sin \theta e^{(F_1 + F_2)|_{r_H}} \ .
\end{equation}
The entropy follows from the Bekenstein-Hawking formula, $S = A_H/4$, and the horizon angular velocity is found evaluating the \textit{ansatz} function $W$ at the event horizon, $\Omega_H = W|_{r_H}$.

At spatial infinity, on the other hand, the ADM mass, $M$, and total angular momentum, $J$, are computed from the asymptotic behaviour of  the metric functions:
\begin{equation}
g_{tt} = -e^{2F_0} N + e^{2F_2} W^2 r^2 \sin^2 \theta \rightarrow -1 + \frac{2M}{r} + \dots \hspace{10pt}, \hspace{10pt} g_{\phi t} = -e^{2F_2} W r^2 \sin^2 \theta  \rightarrow - \frac{2J}{r} \sin^2 \theta + \dots~.
\end{equation}

The above quantities, together with two new ones, are related by a \textit{Smarr-type formula}~\cite{Smarr:1972kt},
\begin{equation}
M = 2 T_H S + 2 \Omega_H (J - mQ) + M^{\Psi} \ ,
\end{equation}
where two new quantities appear: the scalar field energy (mass), $M^{\Psi}$,
\begin{equation}
M^{\Psi} = \int_{\Sigma} dS_a (2 T^a_b \xi^b - T \xi^{a} ) = 4\pi \int^\infty_{r_H} dr \int^\pi_0 d\theta\ r^2 \sin \theta\ e^{F_0 + 2F_1 + F_2} \left[ \mu^2 - 2e^{-2F_2} \frac{\omega (\omega - m W)}{N} \right] \phi^2 \ ,
\end{equation}
(with the Killing vector $\xi=\partial_t$), 
and the Noether charge associated to the global $U(1)$ symmetry of the scalar field, $Q$,
\begin{equation}
Q = 4\pi \int^\infty_0 dr \int^\pi_0 d\theta\ r^2 \sin \theta\ e^{-F_0 + 2F_1 + F_2} \frac{\omega - m W}{N} \phi^2 \ .
\end{equation}
The Noether charge is, moreover, related with the scalar field angular momentum as $J^\Psi = m Q$. This has suggested the definition of a dimensionless parameter that quantifies how hairy a given BH is: 
\begin{equation}
q \equiv  \frac{J^{\Psi}}{J} = \frac{m Q}{J} \ .
\end{equation}
If $q = 0$ the BH has no scalar hair; this is the Kerr BH limit. On the other end of the spectrum, if $q = 1$, all angular momentum is in the scalar hair; in fact, this is no longer a BH but rather an everywhere regular solitonic solution, corresponding to the \textit{boson star} limit. In this case, all angular momentum is quantised in terms of the Noether charge~\cite{Schunck:1996he,Yoshida:1997qf,Kleihaus:2005me}. In between, when $0 < q < 1$, hairy BHs exist.

\section{Domain of Existence}
\label{section3}

\begin{figure}[h!]
\centering
\subfloat[][]{\includegraphics[scale=0.65]{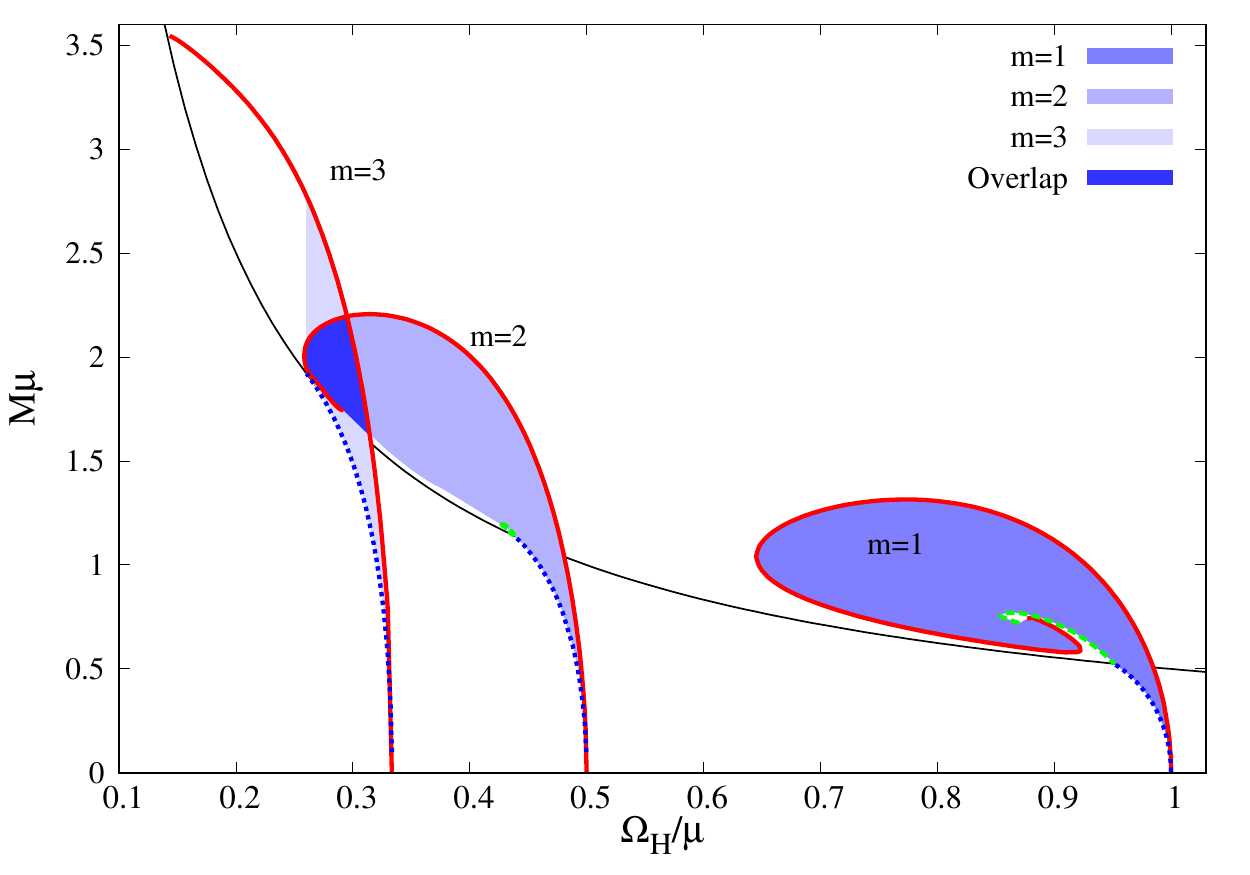}
\label{SubFig:KBHsSH_OmegaH_Mass}}
\subfloat[][]{\includegraphics[scale=0.65]{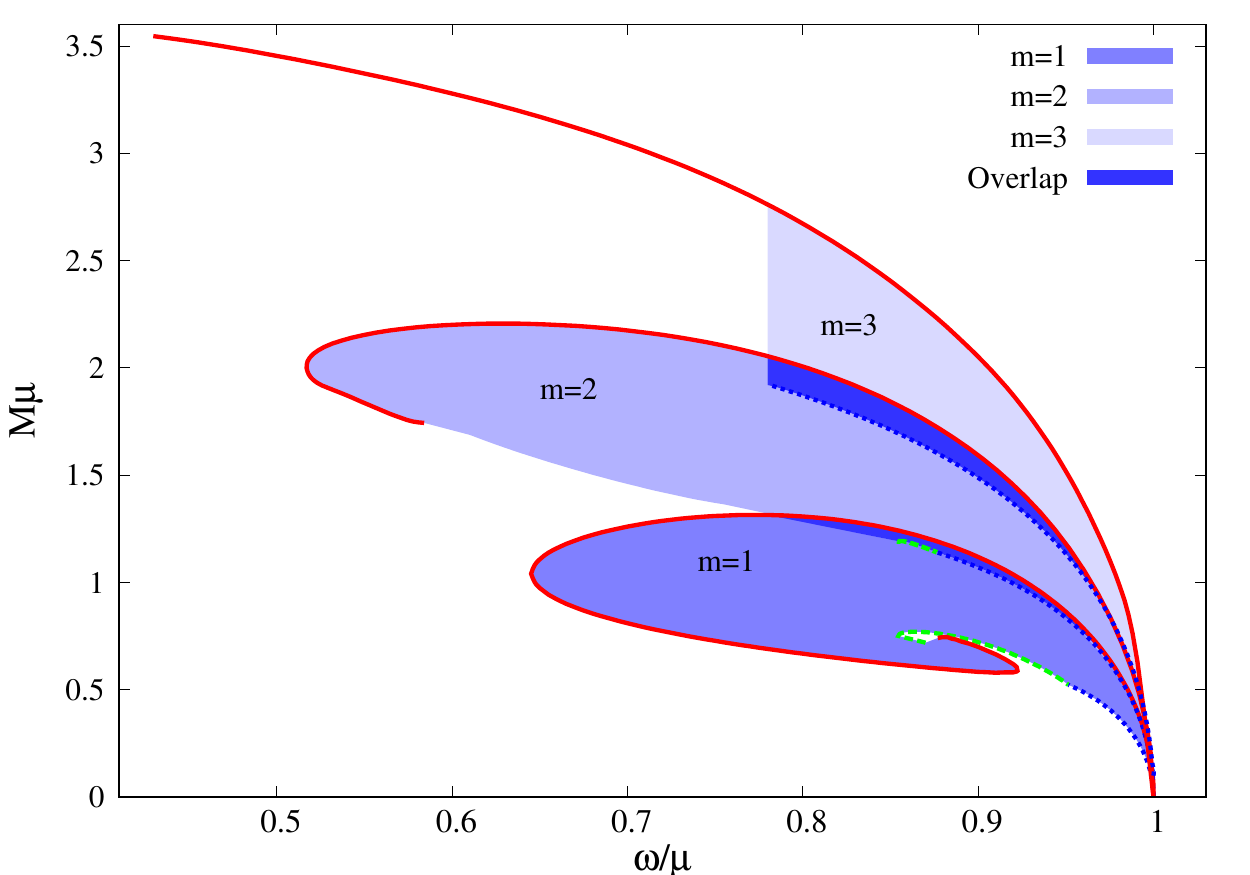}
\label{SubFig:KBHsSH_w_Mass}}
\caption{Domain of existence in an ADM mass $vs.$: \textbf{(a)} Event horizon angular velocity, $\Omega_H$; \textbf{(b)} Scalar field angular frequency, $\omega$. The red line represents the boson star line, the blue dotted line is the existence line, and the green dashed line is the line of extremal hairy BHs. Solutions exist in the domain (shaded blue regions) bounded by these three lines. The black solid line (left panel) describes extremal Kerr BHs: the Kerr family of BHs exists on and below that line. The colour scheme is kept in the subsequent figures.}
\label{Fig:KBHsSH_ParameterSpace1}
\end{figure}

Fixing $n=0$, the domain of existence spanned by the hairy BHs is a 2-dimensional space. In our framework to construct the solutions,
 with dimensionless natural units set by $\mu$~\cite{Herdeiro:2015gia}, 
this domain is scanned by varying the angular frequency of the scalar field, $\omega$, and the radial coordinate of the event horizon, $r_H$.  
Such 2-dimensional region can, however, be exhibited in several more physically meaningful ways, as $r_H$ is not physically meaningful \textit{per se}. In Fig. \ref{Fig:KBHsSH_ParameterSpace1}, 
following previous literature, the domain of existence is shown in an ADM mass \textit{vs.} horizon angular velocity (Fig. \ref{SubFig:KBHsSH_OmegaH_Mass}) and in an ADM mass \textit{vs.} scalar field angular frequency (Fig. \ref{SubFig:KBHsSH_w_Mass}) plots. Both panels exhibit the domain of existence of the hairy BHs with $m=1$, $m=2$ and $m=3$. For the $m=3$ case, only a part of the domain of existence is shown, corresponding to a region of interest for the entropic comparison. The left panel shows, moreover, the region where vacuum Kerr BHs exist -- below the black solid line in Fig. \ref{SubFig:KBHsSH_OmegaH_Mass}.

The domain of existence of the hairy BHs is shown in Fig. \ref{Fig:KBHsSH_ParameterSpace1} as the shaded blue regions, corresponding to the extrapolation to continuum of isolated numerical points. It is bounded by three curves:
\begin{itemize}
\item The \textit{boson star line} - corresponding to the solitonic limit, in which both the event horizon radius and the horizon area vanish, $r_H = 0$ and $A_H = 0$,  and the solution has no BH, thus $q = 1$. Such line is represented in both subfigures in Fig. \ref{Fig:KBHsSH_ParameterSpace1} as a red solid line.
\item The \textit{extremal line} -  corresponding to extremal hairy BHs, which, by definition, have a vanishing Hawking temperature, $T_H  =0$. Such line is represented in both subfigures in Fig. \ref{Fig:KBHsSH_ParameterSpace1} as a green dashed line.
\item The \textit{existence line} - corresponding to specific subset of vacuum Kerr BHs which can support scalar clouds. These solutions have $q = 0$. Such line is represented in both subfigures in Fig. \ref{Fig:KBHsSH_ParameterSpace1} as a blue dotted line.
\end{itemize}

Firstly, consider the right panel (Fig. \ref{SubFig:KBHsSH_w_Mass}). 
As $m$ increases, the domain of existence broadens up in its frequency range, allowing hairy BHs with lower angular frequencies and larger ADM masses.
Each $m$ family can overlap with the previous $m-1$ family, where is possible to have hairy BHs with the same angular frequency and ADM mass but with different $m$. Observe, however, that the regions of overlap for $m=1,2$ and $m=2,3$ solutions are distinct. Thus, three consecutive $m$ families do not overlap.

In Fig. \ref{SubFig:KBHsSH_OmegaH_Mass}, on the other hand, one observes that there is no region of overlapping  $m=1,2$ solutions. 
Two hairy BHs with different  $m$, and $m=1,2$, can have the same ADM mass, but not the same horizon angular velocity.
The same can not be said for $m=2,3$ solutions: there is a region of overlap. Nonetheless, by cross-checking information from Fig. \ref{SubFig:KBHsSH_w_Mass} and Fig. \ref{SubFig:KBHsSH_OmegaH_Mass} one can establish that no two hairy BHs with the same ADM mass, angular frequency, $\omega$, and horizon angular velocity, $\Omega_H$ exist, in the $m=2,3$ overlap. This overlap in Fig. \ref{SubFig:KBHsSH_w_Mass} occurs for large angular frequencies, which correspond to solutions close to $\Omega_H \sim 0.5$; in Fig. \ref{SubFig:KBHsSH_OmegaH_Mass}, for $m=2,3$, on the other hand, one can see that the overlapping solutions occur only around $\Omega_H \sim 0.3$. 

 In Fig. \ref{Fig:KBHsSH_ParameterSpace1} $m=2$ ($m=3$) solutions  have an horizon angular velocity which is half (one third) of the allowed angular frequency -- cf. Fig. \ref{SubFig:KBHsSH_w_Mass}. For $m=1$, the scalar field angular frequency is equal to the horizon angular velocity, and the domain of existence of hairy BHs with $m=1$ is exactly the same in both plots.

\section{Phase space}
\label{section4}

\begin{figure}[h!]
\centering
\subfloat[][]{\includegraphics[scale=0.65]{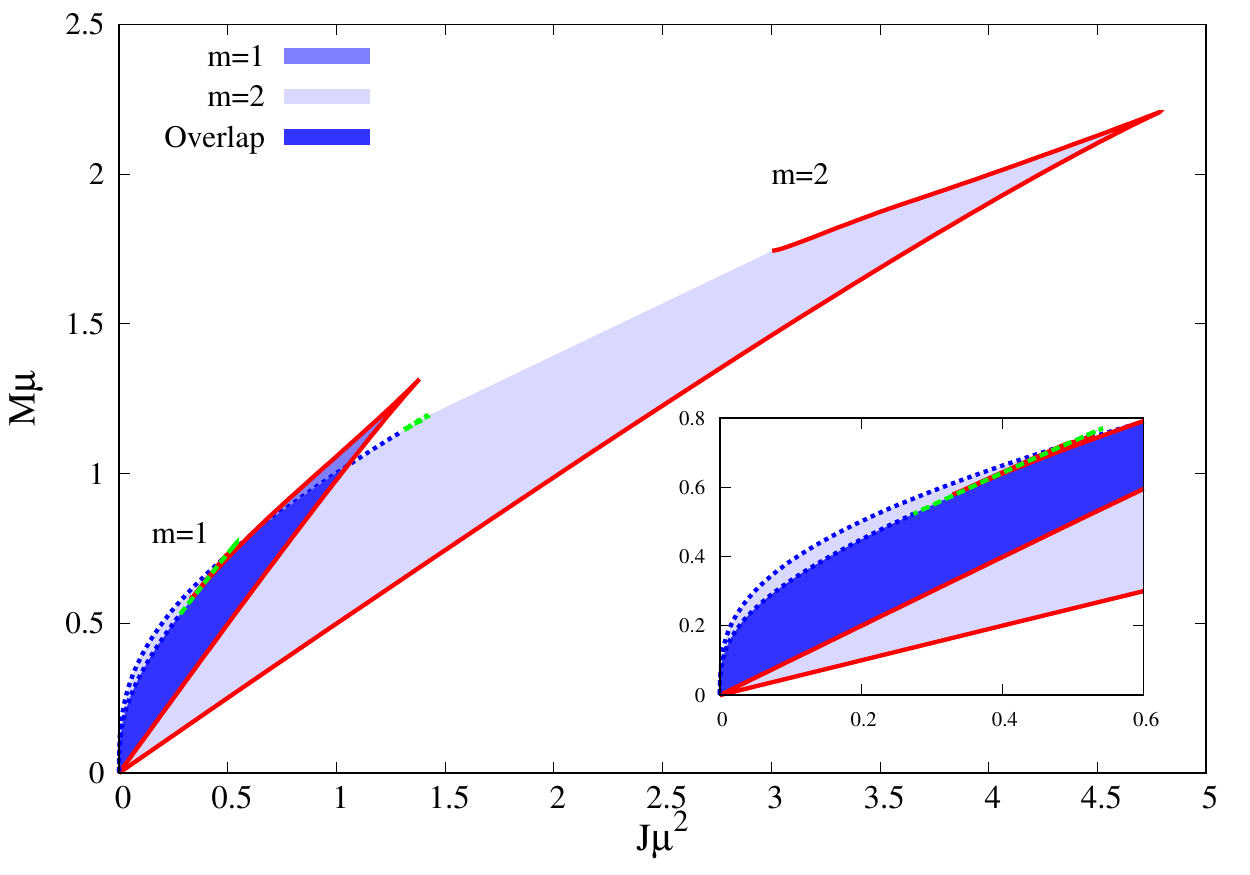}
\label{SubFig:KBHsSH_J_Mass}}
\subfloat[][]{\includegraphics[scale=0.65]{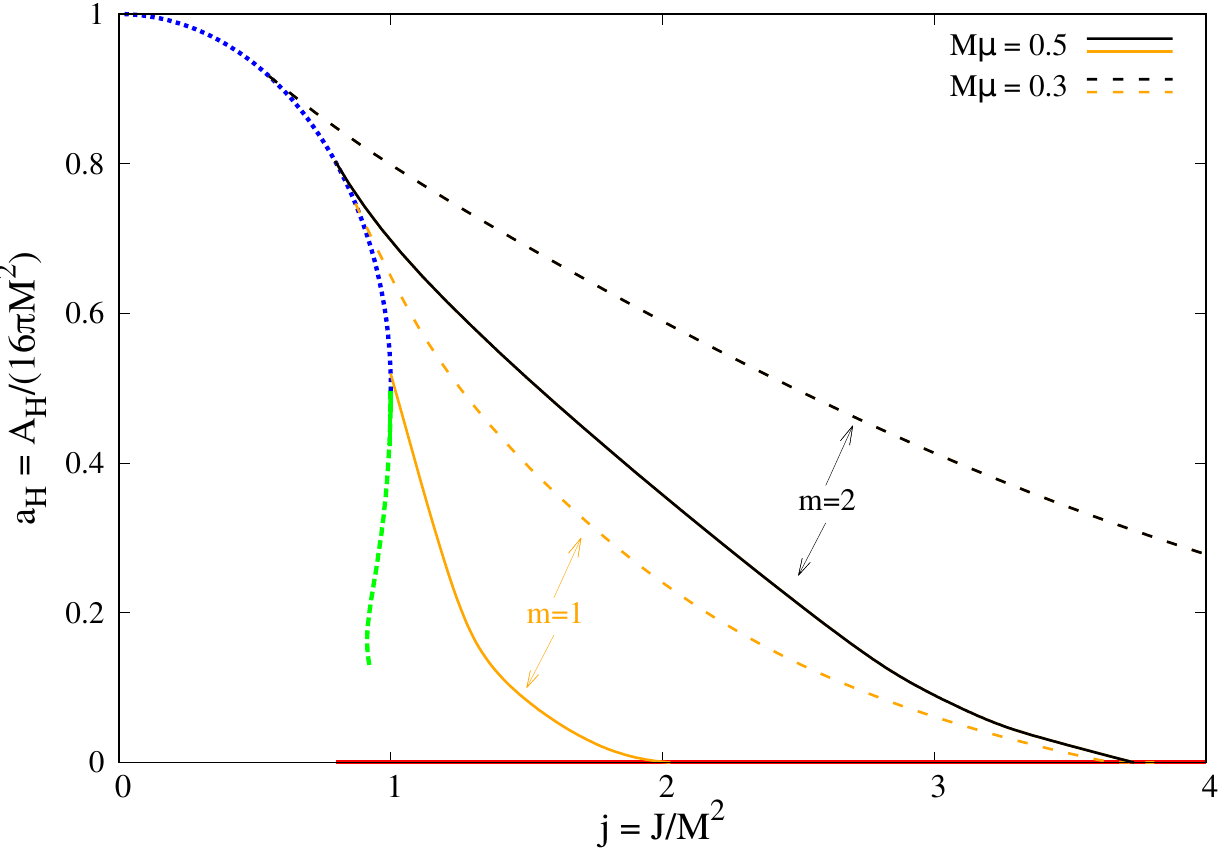}
\label{SubFig:KBHsSH_j_aH}}
\caption{\textbf{(a)} ADM mass $vs.$ total angular momentum for the $m=1$ and $m=2$ families; \textbf{(b)} Reduced horizon area, $a_H$, $vs.$ reduced spin, $j$.  The (orange, for $m=1$ and black, for $m=2$) curves correspond to solutions with constant ADM mass: dashed (solid) lines correspond to $M\mu = 0.3$ ($M\mu = 0.5$).}
\label{Fig:KBHsSH_ParameterSpace2}
\end{figure}

\begin{figure}
\centering
\subfloat[][]{\includegraphics[scale=0.65]{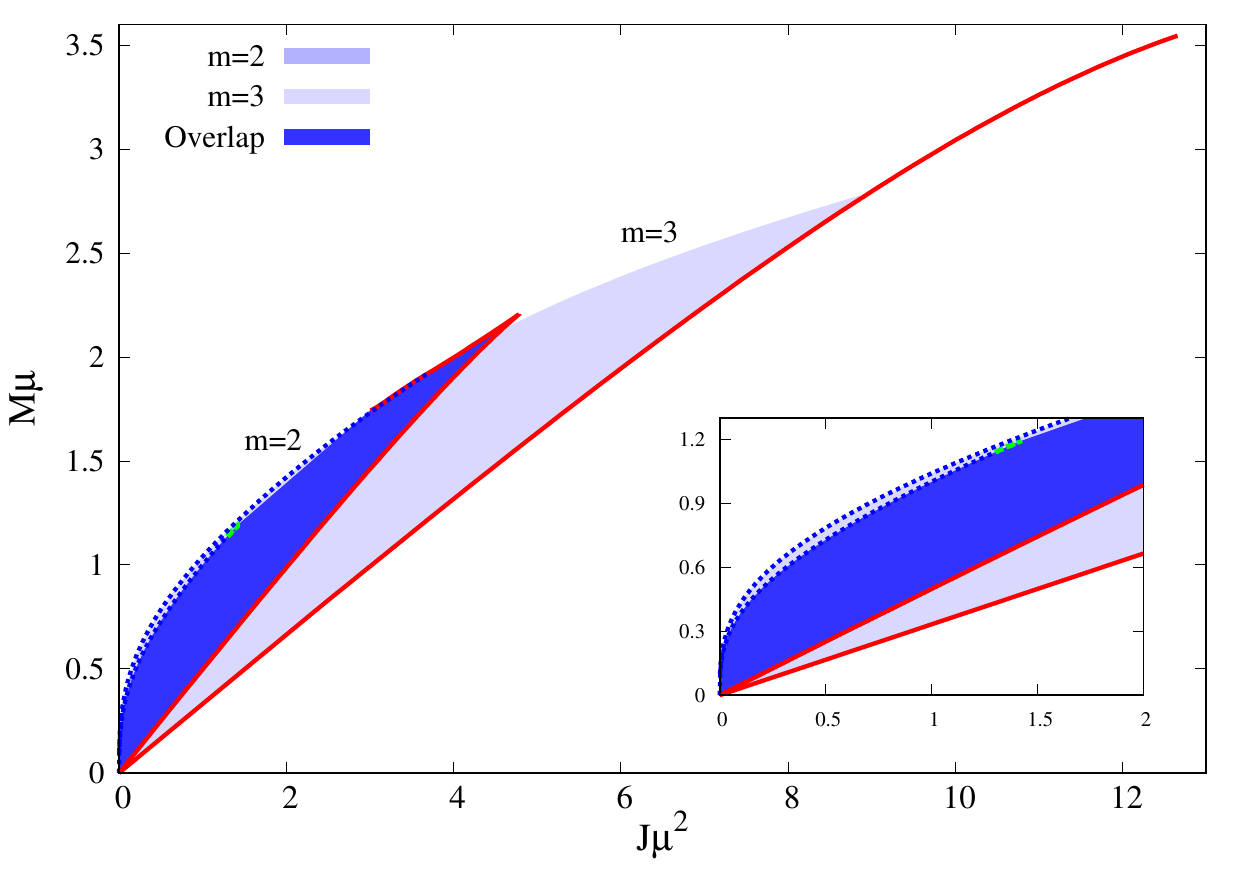}
\label{SubFig:KBHsSH_J_Mass_with_m_3}}
\subfloat[][]{\includegraphics[scale=0.65]{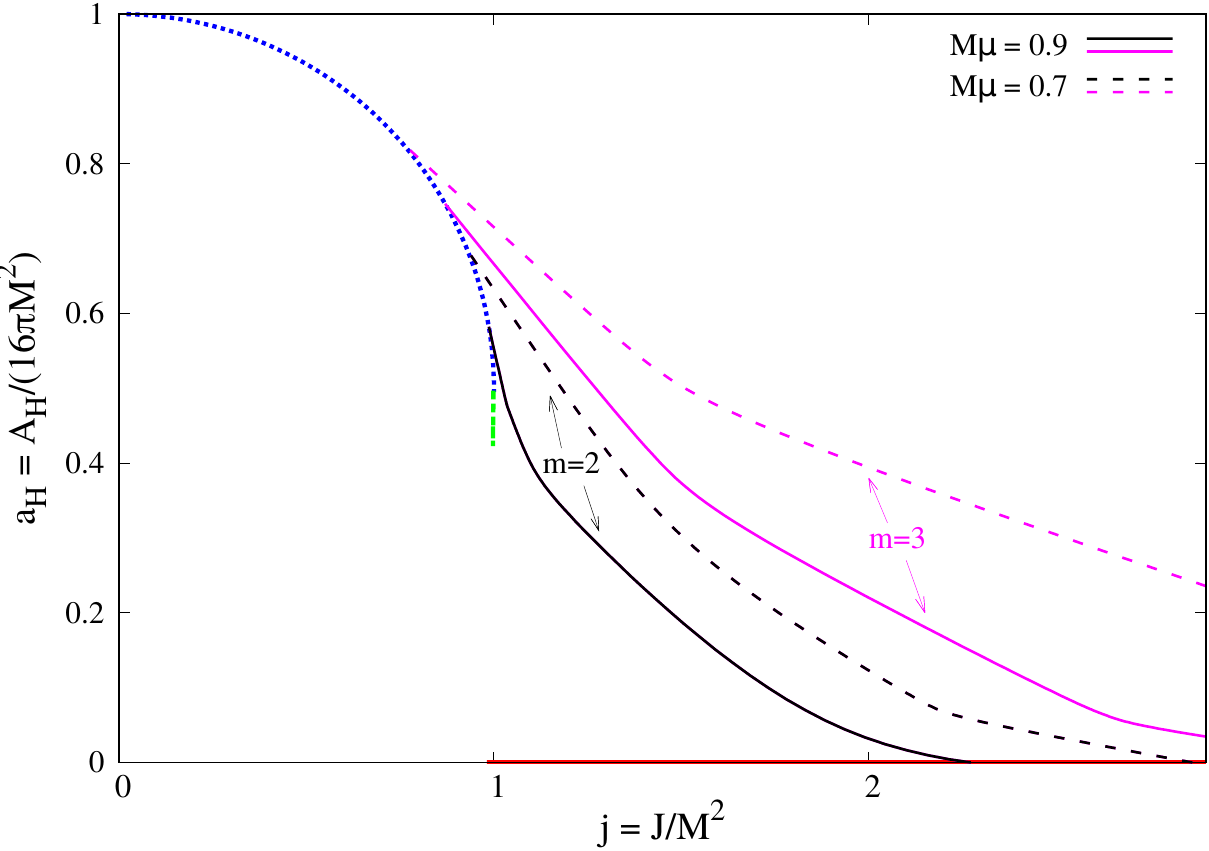}
\label{SubFig:KBHsSH_j_aH_with_m_3}}
\caption{
\textbf{(a)} ADM mass $vs.$ total angular momentum for the $m=2$ and $m=3$ families; \textbf{(b)} Reduced horizon area, $a_H$, $vs.$ reduced spin, $j$.  The (black, for $m=2$ and pink, for $m=3$) curves correspond to solutions with constant ADM mass: dashed (solid) lines correspond to $M\mu = 0.7$ ($M\mu = 0.9$).}
\label{Fig:KBHsSH_ParameterSpace22}
\end{figure}

Let us now analyse the domain of existence in the total $(M,J)$ space, $i.e.$ phase space. This is represented in Fig. \ref{SubFig:KBHsSH_J_Mass} and Fig. \ref{SubFig:KBHsSH_J_Mass_with_m_3}. 
Fig. \ref{Fig:KBHsSH_ParameterSpace1} already made manifest that solutions with higher $m$ are allowed to be more massive; this is confirmed in Figs. \ref{SubFig:KBHsSH_J_Mass} and Fig. \ref{SubFig:KBHsSH_J_Mass_with_m_3}. The latter, moreover, show that higher $m$ solutions can have larger angular momentum, thus broadening the domain of existence. Furthermore, a region of overlapping solutions is again manifest: there are hairy BHs with different $m$ but with the same $(M,J)$. A natural question is then, which amongst these degenerate solutions, in terms of global quantities, is entropically preferred.

In Fig. \ref{SubFig:KBHsSH_j_aH} the reduced horizon area, $a_H \equiv A_H/16\pi M^2$ is shown as a function of the reduced spin, $j \equiv J/M^2$, for hairy BHs belonging to the $m=1,2$ families (orange lines represent $m=1$; black lines represent $m=2$), with two illustrative values for the ADM mass, $M\mu = 0.3$ (dashed lines) and $M\mu = 0.5$ (solid lines). Observe that the existence line (dashed blue line) is common to both families. These lines follow the Kerr relation,
\begin{equation}
a_H^{\text{Kerr}} = \frac{1}{2} \left(1+\sqrt{1-j_{\text{Kerr}}^2} \right) \hspace{2pt} \ . 
\end{equation}
The extremal BH line (dashed green lines) of both $m$ families, on the hand, overlap only at the point wherein they touch the existence line. Beyond this point, both lines are close but do not overlap and most of the green line seen in Fig. \ref{SubFig:KBHsSH_j_aH} corresponds to the $m=1$ solutions. The figure also exhibits two illustrative pairs of lines corresponding to sequences of hairy BHs with the same $M$ (solid black and orange lines for $M\mu = 0.5$, or dashed black and orange line for $M\mu = 0.3$), demonstrating that the solutions with $m=2$ will always have a larger horizon area and hence a larger entropy, when both solutions have the same $j$. A similar analysis is performed in Fig. \ref{SubFig:KBHsSH_j_aH_with_m_3}, for $m=2,3$ solutions with similar conclusions. We remark that in this case only the $m=2$ extremal line is shown, as this line was not computed in the $m=3$ case.

\section{Other physical properties}
\label{section5}
Let us now briefly consider other salient properties of the hairy BHs with $m>1$.

\subsection{Temperature distribution and Kerr bound violation}

\begin{figure}[h!]
\centering
\subfloat[][]{\includegraphics[scale=0.65]{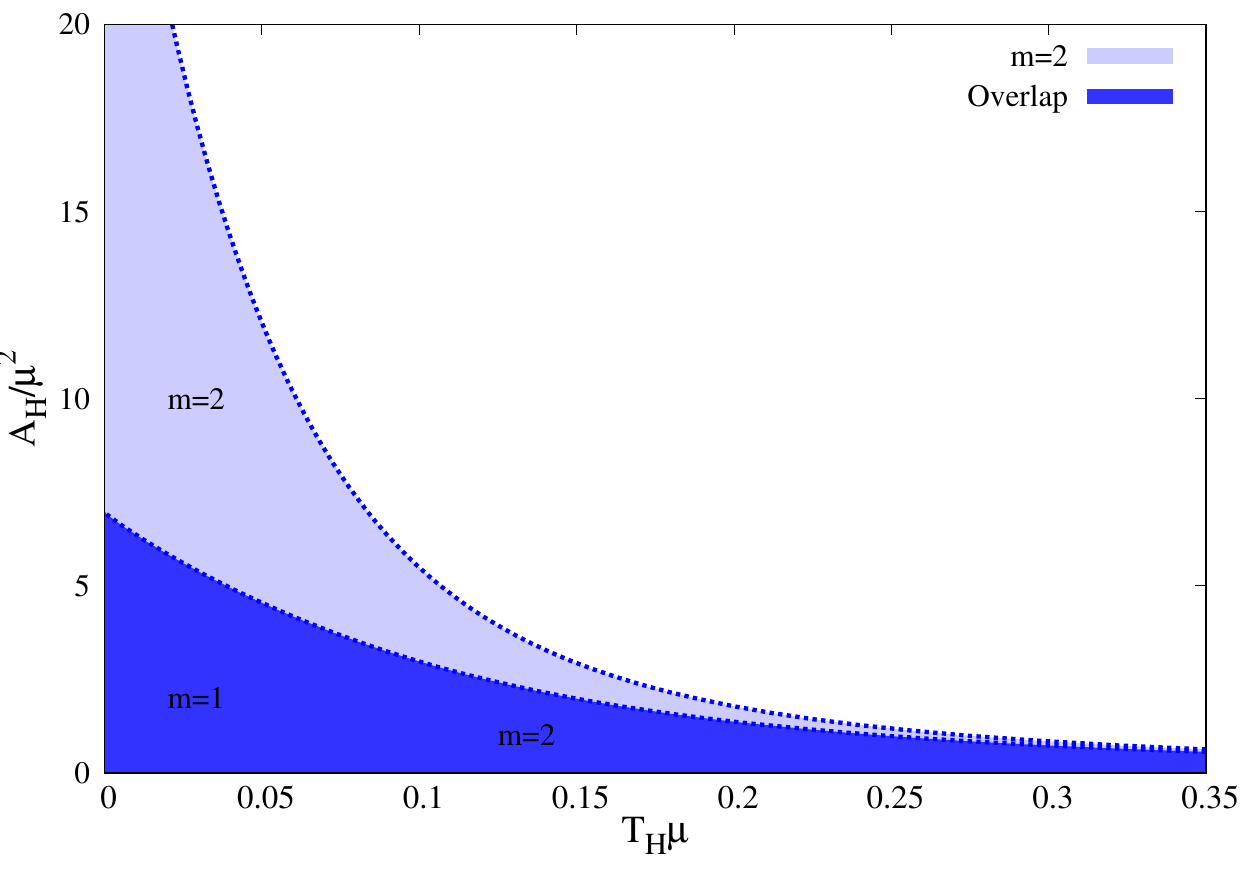}
\label{SubFig:KBHsSH_TH_AH}}
\subfloat[][]{\includegraphics[scale=0.65]{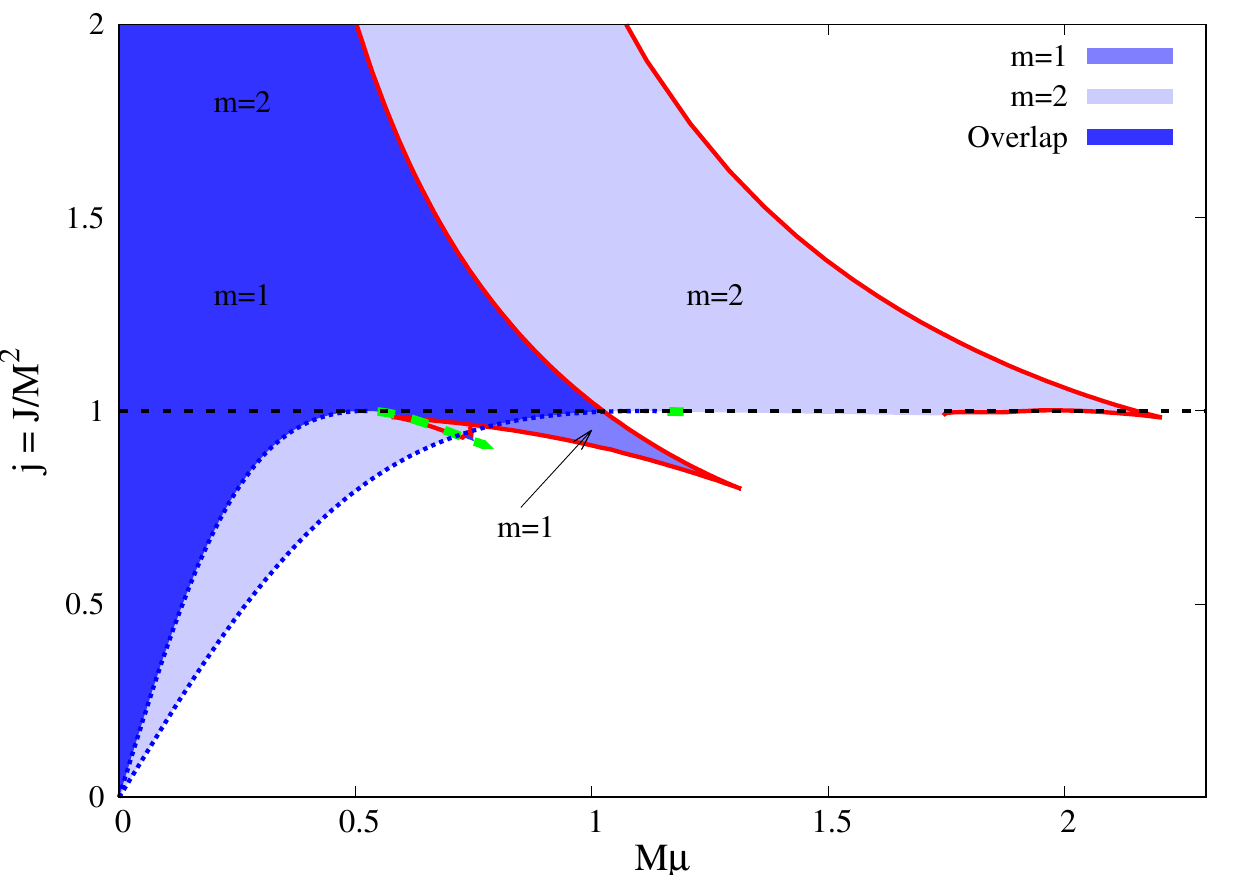}
\label{SubFig:KBHsSH_Mass_j}}
\caption{\textbf{(a)} Horizon area, $A_H$ $vs.$ the Hawking temperature, $T_H$; \textbf{(b)} Reduced spin, $j=J/M^2$ $vs.$ the ADM mass. The black dotted line corresponds to the Kerr bound where $j=1$. Hairy solutions with both $m=1$ and $m=2$, can violate this bound.}
\label{Fig:KBHsSH_ParameterSpace3}
\end{figure}

In Fig. \ref{SubFig:KBHsSH_TH_AH} we exhibit the horizon area, $A_H$ of $m=1,2$ hairy BHs $vs.$ their Hawking temperature, $T_H$.  Fixing $T_H$, there are always hairy BHs with $m=2$ with larger horizon area and hence entropically preferred. Likewise, fixing $A_H$, there are always $m=2$ solution with a larger Hawking temperature than $m=1$ solutions.

In Fig. \ref{SubFig:KBHsSH_Mass_j}, the reduced spin, $j = J/M^2$, is exhibited in terms of the ADM mass of the hairy BHs. This confirms a result already manifest in Fig. \ref{SubFig:KBHsSH_j_aH}. For Kerr BHs there is a limit to the reduced spin they can carry; if a Kerr BH rotates too fast, no event horizon is possible. This is the \textit{Kerr bound}, $j \leqslant 1$. Figs. \ref{SubFig:KBHsSH_j_aH}, \ref{SubFig:KBHsSH_j_aH_with_m_3} and Fig. \ref{SubFig:KBHsSH_Mass_j}, confirm that the existence line (vacuum Kerr BHs that can support scalar clouds) only extends to $j=1$, obeying the Kerr bound, but hairy BHs of both $m$ families, can violate the Kerr bound. In fact, for constant $M$, larger $m$ solutions have stronger violations of the bound.

\subsection{Ergoregions}

Kerr BHs are well known to possess an ergoregion~\cite{Chandrasekhar:1985kt}, wherein the asymptotically timelike Killing vector field becomes spacelike outside the event horizon. In such region, the BH has to perform work on any causally moving object~\cite{Penrose1971}, which by energy conservation means the BH transfers some of its rotational energy to such an object. The existence of an ergoregion is at the source of the Penrose process, superradiant scattering and superradiant instabilities; the latter trigger the migration of the Kerr BH and hairy BH solutions towards higher $m$ in Einstein-(massive, complex-)Klein-Gordon models. 

The typology of ergoregions in the $m=1$ hairy solutions is richer than in Kerr~\cite{Herdeiro:2014jaa}. In the former case, BHs can have two different types of ergoregions: an ergo-sphere -- the same as Kerr BHs; or an ergo-Saturn. The latter is the superposition of the standard BH ergo-sphere and an ergo-torus known to be present in some fast rotating boson stars~\cite{Kleihaus:2007vk}. 

\begin{figure}[h!]
\centering
\includegraphics[scale=0.75]{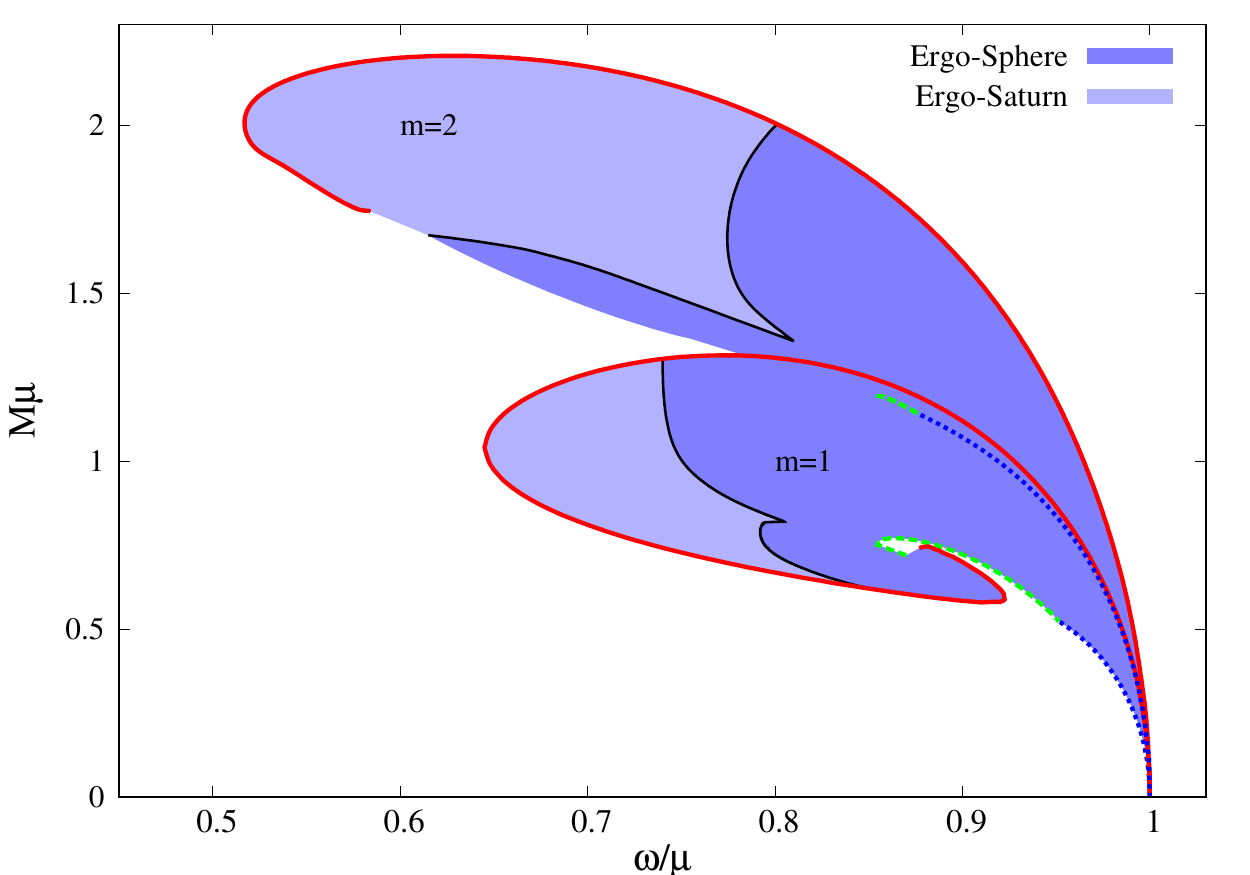}
\caption{Ergo-regions typologies. Hairy BHs develop an ergo-sphere in the dark blue shaded region and an ergo-Saturn in the light blue shaded region.}
\label{Fig:KBHsSH_Ergoregions}
\end{figure}

In Fig. \ref{Fig:KBHsSH_Ergoregions} we show how the  typology of ergoregions is distributed in the domain of existence of hairy BHs with $m=1,2$. The distribution is qualitatively similar in both cases. Ergo-spheres exist in the hairy BHs that connect to boson stars without ergoregions and also in the vicinity of the Kerr limit.  Ergo-Saturns, on the other hand, only exist in the parts of the domain of existence of lower frequency, in the neighbourhood of the boson star solutions that possess an ergo-torus. The transition from solutions that possess only an ergo-sphere to the ones with the composite structure of an ergo-Saturn is similar to that found in the $m=1$ case and which is detailed in Fig. 3 in~\cite{Herdeiro:2014jaa}. Similar ergo-Saturns were recently reported in a different model of BHs with synchronised hair~\cite{Herdeiro:2018djx}.

\subsection{Horizon isometric embedding}

As a final physical aspect let us consider the horizon geometry of the higher $m$ hairy BHs. This can be analysed by isometrically embedding the spatial sections of the horizon in Euclidean 3-space, $\mathbb{E}^3$. We will follow~\cite{Delgado:2018khf} and focus on the $m=2$ solutions. Let us start with a brief summary of the procedure.

From eq. \eqref{Eq:NumericMetric},  the induced metric on the spatial sections of the horizon is,
\begin{equation}\label{Eq:InducedMetric}
d\sigma^2 = r_H^2 \left( e^{2F_1(r_H,\theta)} d\theta^2 + e^{2F_2(r_H,\theta)} \sin^2 \theta d\varphi^2 \right)  \ .
\end{equation}
To embed this 2-surface in $\mathbb{E}^3$, with a Cartesian metric,
\begin{equation}\label{Eq:EuclideanMetric}
d\sigma^2 = dX^2 + dY^2 + dZ^2 \ ,
\end{equation}
one uses the embedding functions, $f(\theta)$ and $g(\theta)$, which make use of the axi-symmetry of the 2-surface,
\begin{equation}
X + i Y = f(\theta) e^{i\varphi} \hspace{5pt} , \hspace{5pt} Z = g(\theta) \ .
\end{equation}
For our case, the embedding functions can be chosen as:
\begin{equation}
f(\theta) = e^{F_2(r_H,\theta)} r_H \sin \theta \hspace{10pt} , \hspace{10pt} g'(\theta) = r_H \sqrt{k(\theta)} \ ,
\end{equation}
in which the function $k(\theta)$ is defined as $k(\theta) = e^{2F_1(r_H,\theta)} - e^{2F_2(r_H,\theta)} \left[ F_2'(r_H,\theta) \sin \theta + \cos \theta \right]^2$, and the prime $'$ denoted the derivative in order to the coordinate $\theta$. 

Following~\cite{Delgado:2018khf}, it is possible to show that  in order to have a global embedding, a necessary and sufficient condition is that  $k(\theta) \geqslant 0, \hspace{2pt} \forall \hspace{4pt} \theta \in [0, \pi]$, and this is assured iff the second derivative of the $k(\theta)$ function evaluated at the poles is non-negative, \textit{i.e.},
\begin{equation}
k''(0) \geqslant 0 \ .
\end{equation}
Solving this inequality yields,
\begin{equation}\label{Eq:EmbeddabilityCondition}
F_1''(r_H,0) - 3F_2''(r_H,0) + 1 \geqslant 0 \ .
\end{equation}
On the other hand, the Gaussian curvature of the horizon at the poles is given by,
\begin{equation}
\mathcal{K}|_{\theta = \{0, \pi \}} = \frac{e^{-2F_1(r_H,0)}}{r_H^2} \left[ F_1''(r_H,0) - 3F_2''(r_H,0) + 1 \right] \ .
\end{equation}
Thus a necessary and sufficient condition for a global embedding in $\mathbb{E}^3$ to exist is that the curvature at the poles is non-negative. This conclusion was first obtained in the Kerr-Newman case by Smarr~\cite{Smarr:1973zz}. Thus, the threshold of embeddability occurs when the Gaussian curvature vanishes at the poles. The sequence of hairy solutions that occur at this threshold composed the~\textit{Smarr Line}~\cite{Delgado:2018khf}.

\begin{figure}[h!]
\centering
\includegraphics[scale=0.75]{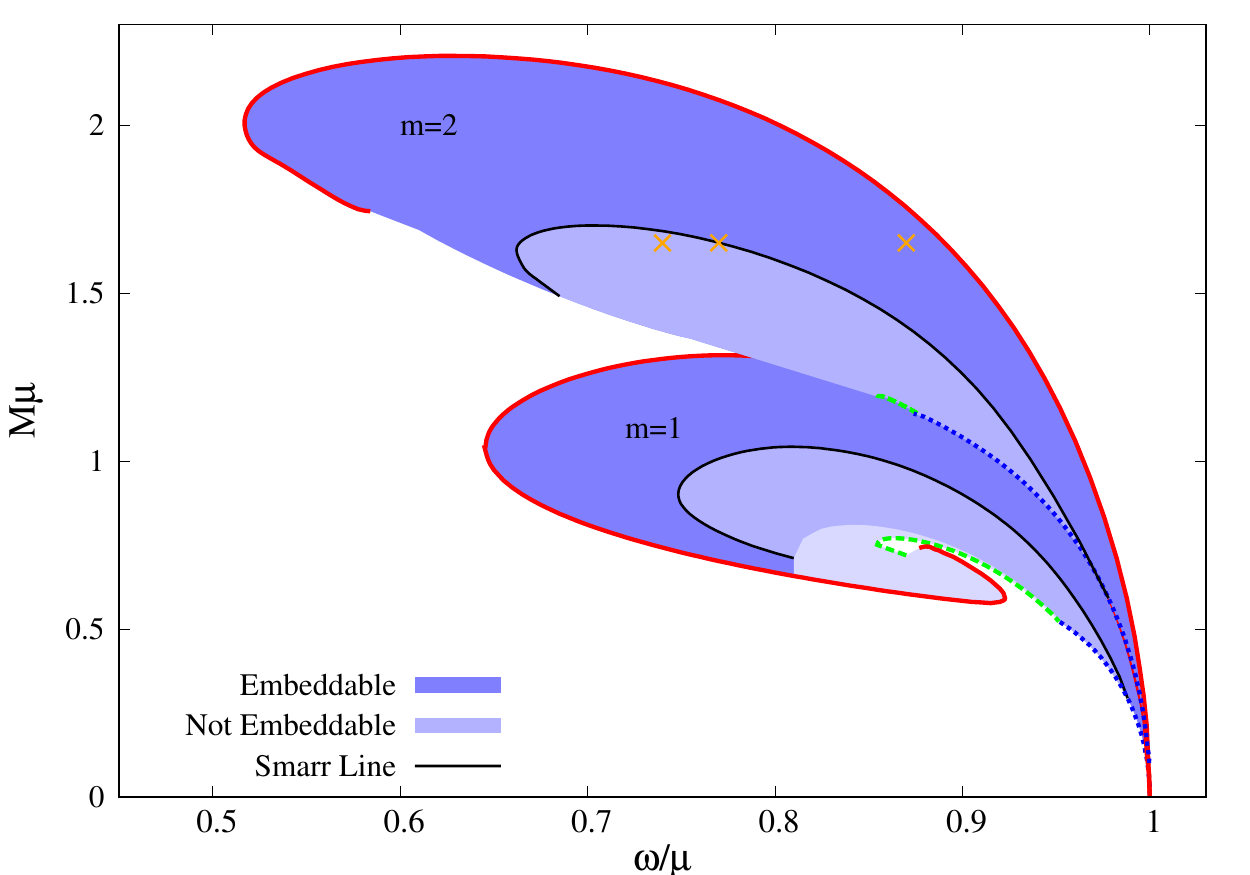}
\caption{Smarr line in the domain of existence, for both $m=1,2$ families. The dark (medium) blue regions correspond to embeddable (non-embeddable) solutions. The light blue region for $m=1$ solutions corresponds to solutions that were not analysed due to numerical accuracy. Three illustrative solutions are highlighted (with crosses) all with the same ADM mass, $M\mu = 1.65$.}
\label{SubFig:KBHsSH_w_Mass_SmarrLine}
\end{figure}

In Fig. \ref{SubFig:KBHsSH_w_Mass_SmarrLine} we present the Smarr (black solid) line in the domain of existence of both $m=1,2$ solutions.
The Smarr line divides the domain of existence into the embeddable region (medium blue) and non-embeddable region (dark blue). This division is qualitatively similar for both the $m=1,2$ families.  Both Smarr lines start at the existence line, in the exact point where the Kerr BH is no longer embeddable, and both have an inspiral behaviour, attaining first a maximum value of the ADM mass, then a minimum value of the angular frequency of the scalar field and backbends into the opposite direction.

\begin{figure}[h!]
\centering
\includegraphics[scale=0.2]{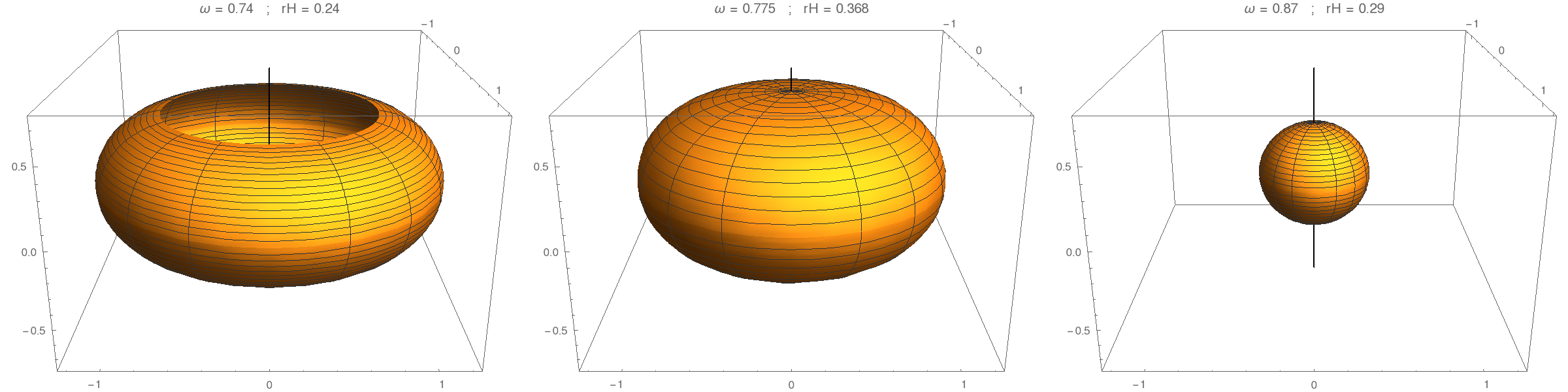}
\caption{Isometric embedding in $\mathbb{E}^3$ of the three highlighted solution in Fig.~\ref{SubFig:KBHsSH_w_Mass_SmarrLine}.  Left panel: Non-embeddable solution; Middle panel: Solution on the Smarr Line; Right panel: Embeddable solution.}
\label{Fig:KBHsSH_ThreeEmbeddings}
\end{figure}

A visualisation of the isometric embedding of the horizon in $\mathbb{E}^3$ is shown in Fig.~\ref{Fig:KBHsSH_ThreeEmbeddings} for the  three hairy BH solutions with the same ADM mass, $M\mu = 1.65$, highlighted in Fig.  \ref{SubFig:KBHsSH_w_Mass_SmarrLine}.
The first solution is within the non-embeddable region, so the embedding misses the region close to the poles. 
The second solution is on the Smarr line, thus this solution will have a zero Gaussian curvature at the poles, therefore such region will appear flat.  The third and final solution is within the embedding region, so it will be possible to draw completely the horizon.  Concerning the latter, we remark that, as this solution is close to the boson star line, where the solutions have a vanishing horizon area, $A_H \rightarrow 0$, its horizon is smaller than that of the previous two solutions. 

\subsection{Horizon sphericity and linear velocity}

To conclude the horizon analysis, following~\cite{Delgado:2018khf} we consider the sphericity, $\mathfrak{s}$, and the horizon linear velocity, $v_H$, in order to assess what is the key property to determine the global embeddability of the horizon. The sphericity measures the deformation of a $U(1)$ invariant compact and simply connected 2-surface when compared to a round sphere, and is defined as,
\begin{equation}
\mathfrak{s} = \frac{L_e}{L_p} \ ,
\end{equation}  
where $L_e$ and $L_p$ are the proper length of the horizon measured around the equator and the poles, respectively. For the hairy BHs it amounts to,
\begin{equation}
\mathfrak{s} = \frac{\pi\ e^{F_2(r_H,\pi/2)}}{\int_0^\pi d\theta e^{F_1(r_H,\theta)}} \ .
\end{equation} 

The horizon linear velocity~\cite{Herdeiro:2015moa} measures how fast the null geodesics generators of the horizon rotate relatively to a static observer at spatial infinity and is defined as,
\begin{equation}
v_H = R \Omega_H \ ,
\end{equation}
where $R \equiv L_e/2\pi$ is the perimetral radius of the circumference at the equator.
For a hairy BH,
\begin{equation}
v_H = e^{F_2(r_H,\pi/2)}r_H \Omega_H \ .
\end{equation}

\begin{figure}
\centering
\subfloat[][]{\includegraphics[scale=0.65]{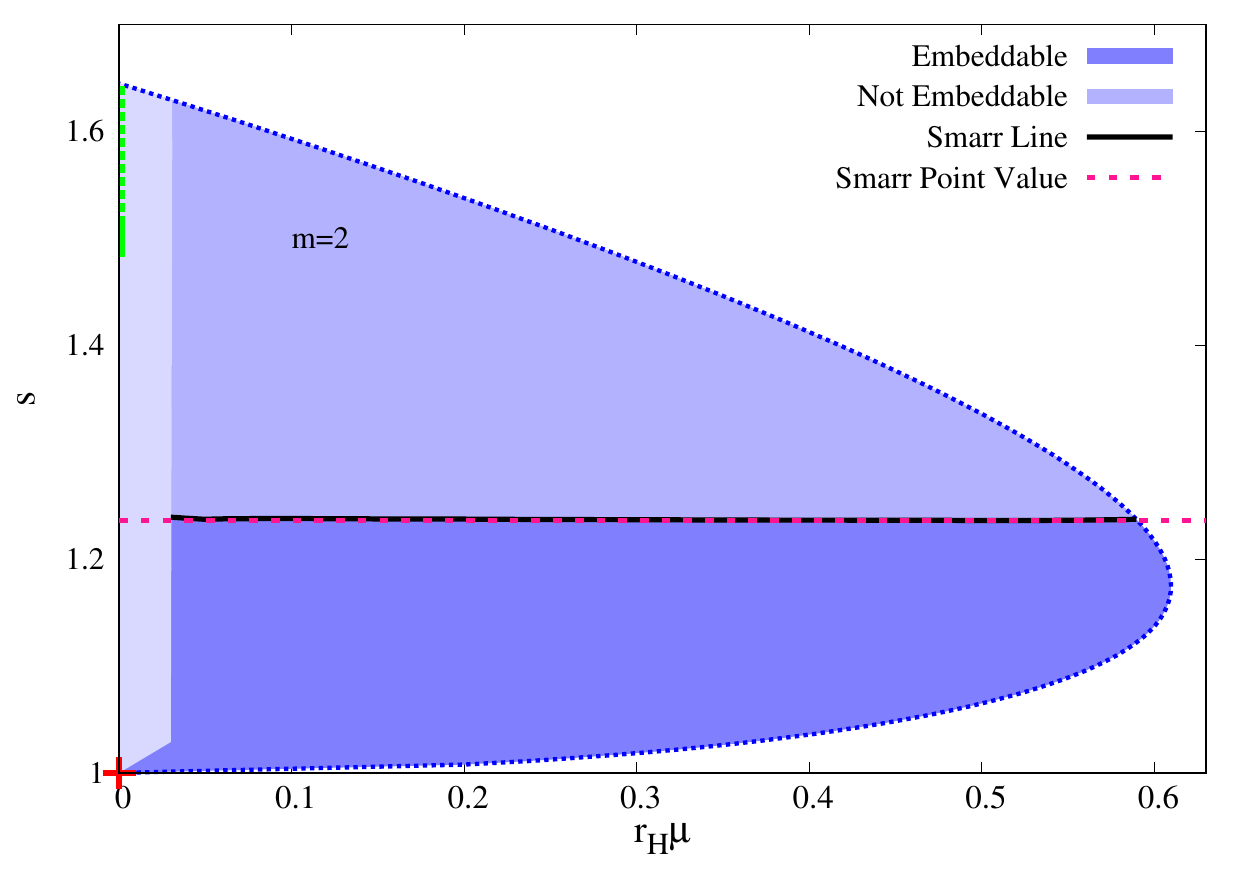}
\label{SubFig:KBHsSH_rH_s}}
\subfloat[][]{\includegraphics[scale=0.65]{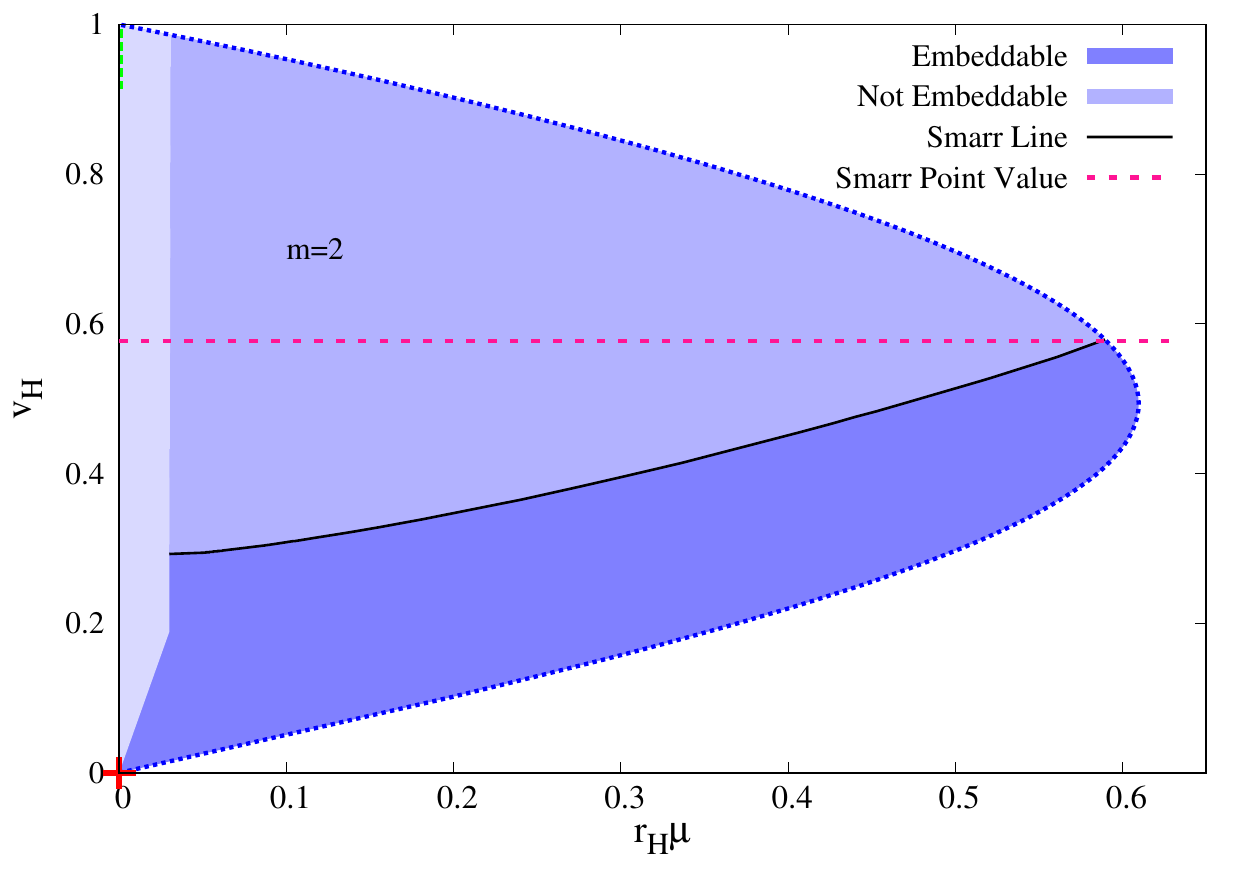}
\label{SubFig:KBHsSH_rH_vH}}
\caption{\textbf{(a)} Sphericity, $\mathfrak{s}$, and \textbf{(b)} horizon linear velocity, $v_H$, $vs.$ the radial coordinate of the event horizon, $r_H$. As before, the dark blue region corresponds to embeddable solutions, and the medium blue represent non-embeddable solutions. The new region of light blue corresponds to solutions that were not analysed due to numerical accuracy.}
\label{Fig:KBHsSH_SmarrLine_PhysicalQuantities}
\end{figure}

Both quantities are exhibited in Fig. \ref{Fig:KBHsSH_SmarrLine_PhysicalQuantities} as a function of the radial coordinate of the horizon, $r_H$. In this representation all hairy solutions are enclosed by the existence line and the vertical line $r_H = 0$, which correspond to both the extremal line -- green dashed line -- and the boson star line -- red cross. The Smarr line is also plotted in both figures, as well as its Kerr limit (the Smarr point). The value at the Smarr point is then extrapolated as a benchmark -- Smarr point value (dashed pink) line. For the sphericity, the Smarr point has a value of $\mathfrak{s}^{(\text{S})} = \frac{\pi}{\sqrt{3}} \frac{1}{E(1/4)} \approx 1.23601$, where $E(k)$ is the complete elliptic integral of second kind, $E(k) =  \int_0^{\pi/2} d\theta \sqrt{1 - k \sin^2 \theta}$; and for the horizon linear velocity, the Smarr point has a value of $v_H^{(\text{S})} = \frac{1}{\sqrt{3}}\approx 0.57735$.

Consider first Fig. \ref{SubFig:KBHsSH_rH_s}. We see that the Smarr line has the same value of sphericity as the Smarr point, within numerical accuracy. Therefore, the sphericity is a faithful diagnosis for embeddability also for $m=2$ solutions: if $\mathfrak{s}$ is lower or equal than $\mathfrak{s}^{(\text{S})}$ than the hairy BH will be embeddable; otherwise, it will be non-embeddable. The same was seen for hairy BHs with $m=1$. 

Now consider Fig. \ref{SubFig:KBHsSH_rH_vH}. None of the hairy solutions exceeds $v_H = 1$, $i.e$ the speed of light. The limit of $v_H = 1$ is only attained by the extremal vacuum Kerr BH. Concerning the Smarr line, unlike what we saw in Fig. \ref{SubFig:KBHsSH_rH_s}, the Smarr line only matches the Smarr point value at in the Kerr limit. The remaining Smarr line solutions will always have a lower $v_H$ than the one obtained at the Smarr point, $v_H^\text{(S)}$. Thus, the value of horizon linear velocity of the Smarr point -- pink dashed line on Fig. \ref{SubFig:KBHsSH_rH_vH} -- is an upper bound, above which all hairy solution with $m=2$ are non-embeddable. Below that bound, both embeddable and non-embeddable solutions exist. The same results were found for $m=1$.

\section{Discussion}
\label{section6}

In this paper, we have constructed and analysed Kerr BHs with synchronised hair and higher azimuthal harmonic index $m$. Specifically, solutions with $m=2,3$ were constructed and contrasted with the $m=1$ solutions. 

There are two results from the analysis that should be emphasised. Firstly, consecutive $m$ families can have degenerate solutions in terms of the global quantities $(M,J)$. When this occurs, the higher $m$ solutions are entropically favoured. This supports the possibility that migrations between such families, triggered by the superradiant instability, could be approximately conservative. This possibility, however, is by no means guaranteed to occur dynamically, as significant gravitational radiation and scalar ejection could take place in this migration. Secondly, there is a high degree of universality in all physical properties that have been unveiled for $m=1$ solutions, that our analysis shows extend \textit{mutatis mutandis} for the higher $m$ solutions. These properties include, in particular, the typology of ergo-regions and the event horizon geometry. There is no reason to expect new qualitative features concerning these physical properties would emerge for even higher $m$ values. 

Similar solutions will also exist in other models of BHs with synchronised hair, for instance including self-interactions~\cite{Herdeiro:2015tia}, or with a Proca field~\cite{Herdeiro:2016tmi}. The results herein indicate no significant differences are to be expected with respect to the $m=1$ case in these models. It would, nonetheless, be interesting to study some phenomenological properties of this higher $m$ solutions, such as BH shadows~\cite{Cunha:2015yba,Cunha:2016bpi}, $X$-ray spectrum~\cite{Ni:2016rhz}, accretion disk morphology~\cite{Gimeno-Soler:2018pjd} or star trajectories~\cite{Franchini:2016yvq}, since these higher $m$ solutions could play a role in the dynamical evolution of the BH/fundamental field system, in case such fundamental, ultra-light, scalar or vector fields exist in Nature.

\section*{Acknowledgements}
This work is supported by the Funda\c{c}\~ao para a Ci\^encia e a Tecnologia (FCT) project UID/MAT/04106/2019 (CIDMA), by CENTRA (FCT) strategic project UID/FIS/00099/2013, by national funds (OE), through FCT, I.P., in the scope of the framework contract foreseen in the numbers 4, 5 and 6
of the article 23, of the Decree-Law 57/2016, of August 29,
changed by Law 57/2017, of July 19. We acknowledge support  from the project PTDC/FIS-OUT/28407/2017 and J. Delgado is supported by the FCT grant SFRH/BD/130784/2017.   This work has further been supported by  the  European  Union's  Horizon  2020  research  and  innovation  (RISE) programmes H2020-MSCA-RISE-2015
Grant No.~StronGrHEP-690904 and H2020-MSCA-RISE-2017 Grant No.~FunFiCO-777740. The authors would like to acknowledge networking support by the
COST Action CA16104.

\bigskip

\bibliography{bibKBHsSH}
\bibliographystyle{ieeetr}

\end{document}